\documentclass[12pt, fleqn]{article}
\usepackage[cp1251]{inputenc}
\usepackage{latexsym,amsfonts,amssymb}
\usepackage{graphicx}
\usepackage{amsbsy}
\usepackage{amsmath}
\usepackage{epsf}

\newcommand{\be} {\begin{equation}}
\newcommand{\ee} {\end{equation}}
\newcommand{\ba} {\begin{array}}
\newcommand{\ea} {\end{array}}
\newcommand{\ts}{\textstyle}  %@@

\newtheorem{theo} {Theorem}% [] %chapter]  %@@
\newcommand{\bteo}{\begin{theo}}  %@@
\newcommand{\et}{\end{theo}}

\newcommand{\pt} {\partial}
\newcommand{\al} {\alpha}
\newcommand{\om} {\omega}
\newcommand{\la} {\lambda}
\newcommand{\ga} {\gamma}
\newcommand{\ta} {\theta}

\begin{document}
\allowdisplaybreaks

\begin{center}
{\Large \bf Lie Symmetries and  Exact Solutions   \\  of  the
Generalized Thin Film Equation  }\\
\medskip

{\bf Roman CHERNIHA~$^{a}$, Phil BROADBRIDGE~$^{b}$ and Liliia MYRONIUK~$^{c}$  } \\
%% \footnote{\small e-mail: cherniha@imath.kiev.ua }\\
\medskip

 {\it $^{a}$~Institute of Mathematics,  NAS
of Ukraine,\\ 3 Tereshchenkivs'ka Street, 01601 Kyiv, Ukraine}\\
  E-mail: cherniha@imath.kiev.ua \\
  {\it $^{b}$~School of Engineering and Mathematical Sciences, \\
  La Trobe University, Bundoora VIC 3083, Australia}\\
  E-mail: P.Broadbridge@latrobe.edu.au\\
{\it $^{c}$~Faculty of  Mathematics,\\
Lesya Ukra\"inka Volyn National University,\\
13 Voly Avenue,  43025 Lutsk, Ukraine} \\
  E-mail: Liliia\underline{  }Myroniuk@univer.lutsk.ua\\

\medskip

\textbf{Abstract}
\end{center}

\medskip

  A symmetry group classification for fourth-order
reaction-diffusion equations, allowing for both second-order and
fourth-order diffusion terms, is carried out. The fourth order
equations are treated, firstly,  as  systems of second-order
equations  that bear some resemblance to systems of coupled
reaction-diffusion equations with cross diffusion, secondly, as
systems of a second-order equation and two first-order equations.
The  paper generalizes  the results of Lie symmetry analysis derived
earlier for particular cases of these equations. Various
 exact solutions are constructed  using Lie symmetry reductions of the
 reaction-diffusion systems
 to ordinary  differential equations.
  The solutions include some unusual structures as well as the familiar types that regularly occur
  in symmetry reductions, namely self-similar solutions, decelerating and decaying traveling waves,
  and steady states.

 \begin{center}
{\textbf{1. Introduction}}
\end{center}

We consider the  fourth order nonlinear partial differential
equation (PDE) of the form
 \be\label{1-1} u_t = -[K(u)u_{xxx}]_x+[D(u)u_x]_x+F(u), \ee
where   $K, \, D$ and $F$ are arbitrary smooth functions ( hereafter
the subscripts $t$ and $x$ denote differentiation with respect to
these variables). Eq. (\ref{1-1}) generalizes a wide range of the
known scalar reaction-diffusion equations arising in applications.
The case with $K$ identically zero and $D(u)>0$ for almost all $u$,
is the  case of second-order reaction-diffusion which has already
been widely studied in  many practical contexts including
combustion, population dynamics, population genetics, neurobiology,
biological cellular growth and adsorptive porous media, e.g.
\cite{Brown}-\cite{Landman}. Hereinafter we assume that $K(u)$ is
not identically zero, so that the governing equation is of the
fourth order, including a fourth-order diffusion term when K is
non-negative. The simplest equation of the form (\ref{1-1}), with
$F=0$ and $D=0$, follows from the approximations of lubrication
theory to describe  thin films of a Newtonian liquid dominated by
surface tension effects. The thin film equations are an active area
of research (see \cite{bertozzi-98}--\cite{king-et-all-07} and
papers cited therein). The equation
 \be\label{1-3} u_t = -[u^\ga u_{xxx}]_x \ee
with the non-negative parameter $\ga$ was introduced in
\cite{greenspan-1978}. $\gamma=3$ describes a classical thin film of Newtonian fluid, as reviewed in \cite{Oron}. $\gamma=1$ occurs in the dynamics of a Hele-Shaw cell \cite{Dupont} and $\gamma=2$ arises in a study of wetting films with a free contact line between film and substrate \cite{bertozzi-98}.

One important generalization of  Eq.
(\ref{1-3}), which is also a particular case  of   Eq.
(\ref{1-1}), reads as \be\label{1-4} u_t = -[u^\ga u_{xxx}]_x+
[u^\mu u_x]_x, \ee where $\mu$ is a positive parameter (arbitrary positive coefficients of each term can be set to one by rescaling variables).
Eq.  (\ref{1-4}) with $\ga=0$  can be considered as a semilinear
limit of the classical Cahn-Hilliard model  of phase
separation \cite{CH}, which is also widely studied (see \cite {cohen-99} and
the papers cited therein). The linear case $\gamma=\mu=0$ also follows from a small-slope approximation to metal surface evolution, with surface-diffusion and evaporation-condensation represented by fourth-order and second-order diffusion terms \cite{Mullins, Cahn}. The case $\gamma=\mu=3$ arises in a study of capillary instability of axisymmetric thin films \cite{Yarin}.

Several papers are devoted to the construction of exact solutions
of the thin film equations (\ref{1-3}) by Lie symmetry reductions
and their  generalizations \cite{Smyth}--\cite{gan-ibr-2008}, or
by searching for special invariant finite vector spaces of
solutions \cite{gal-svi-07}. The symmetry classification is
extended here to include a reaction term. In some circumstances,
the reaction term $F $ should naturally arise in fourth-order
transport equations with a role similar to that in second-order
reaction-diffusion. For example, a particular case of Eq.
(\ref{1-1}) with $F(u)=u$ occurs as the limiting case of the
unstable Cahn-Hilliard equation \cite {segel-84,evans-gal-06}. In
fabricated metal surface evolution, a positive source term may
represent ion beam sputtering \cite{Valbusa} and a negative source
term may represent chemical decay or evaporation \cite{Gaskell}.
Other examples of equations of the form (\ref{1-1}) arising in
applications and having  a reaction term, are presented in
\cite{evans-gal-06}.

The first aim of this paper is to describe all possible Lie symmetries, which  Eq.  (\ref{1-1})
 can admit
depending on the function triplets $(K, D, F)$, i.e. to solve the
so-called group classification problem, which was formulated and
solved for a class of non-linear heat equations in the pioneering
work in \cite{ovs-59}. {This problem for the
 second-order reaction-diffusion equation was solved in \cite{dor}
 (see  also \cite {ch-se-06,ch-se-ra-08} where the problem is solved for the general reaction-diffusion-convection
 equation).
 Note that the most general results concerning non-classical ($Q$-conditional) symmetries
 of reaction-diffusion equations were obtained in  \cite
 {Arrigo}--\cite{a-h}.}

It should be noted that we shall
%%use a non-trivial preliminar step:
not directly
search for Lie symmetries of Eq.  (\ref{1-1}) but
replace one scalar equation by an equivalent cross-diffusion system of equations.
 Using the symmetries found, we construct exact solutions
of  Eq.  (\ref{1-1}) with such   triplets $(K, D, F)$, which
arise in applications and compare the results obtained with those
derived earlier.

%%%We  remind that
 Ovsiannikov's method of Lie symmetry classification of differential equations
  \cite{ovs}
  is based on the classical Lie scheme and a set of
  equivalence transformations
 of a given equation. The formal application of this method to equations containing several arbitrary functions
 (Eq.(\ref{1-1}) contains three arbitrary functions)
 usually leads to a large  number of  equations admitting non-trivial  Lie algebras of invariance.
 Our approach  of Lie symmetry classification
 of differential equations
 is based on the classical Lie scheme and on
 finding and then making  systematic use of the sets of local
%%substitutions
transformations  that reduce any differential equation  with a
Lie algebra of invariance,   to one given in  the relevant  list, that is representative of each equivalence class.
This approach has
earlier been  applied also for reaction-diffusion systems
  \cite{ch-king}--\cite{ch-king5}.

The paper is organized as follows.
 Section 2  is  devoted to a complete description of  Lie symmetries
of Eq. (\ref{1-1}), i.e.  all possible Lie symmetries, which this
equation
  can admit depending on the form the functions $K, \, D$ and $F$, are found.
   In most applications, $K$ and $D$ are non-negative so that the diffusive transport
   processes are dissipative \cite{PB}. { However, in solving a group classification problem,
   it is more common
    to allow the coefficient functions
    to be arbitrary smooth functions.}
  In Section 3, the symmetry reductions and some exact solutions are constructed
  for particular cases of Eq. (\ref{1-1}) that are likely to be useful in applications. The main results of the paper are summarized in the last section.

  \begin{center}
{\textbf{2. Lie symmetry of Eq. (\ref{1-1})}}
\end{center}

 Firstly, we note that Eq. (\ref{1-1}) can be reduced to the system
\be\label{2-1} \ba{l} \medskip
u_t = -[K(u)v_x]_x+[D(u)u_x]_x+F(u),\\
0= u_{xx}-v \ea \ee by the substitution $v=u_{xx}, \, v=v(t,x)$.
Since $u_x$ is not relabeled as a separate variable, point
symmetries of this system do not include contact symmetries of the
original single equation (\ref{1-1}). Since second-order contact
symmetries do not exist \cite{Ibragimov}, point symmetries of this
system could include nothing more than extensions (prolongations) of
point symmetries of the single fourth order equation for $u(x,t)$.
However,  it is convenient to analyse system (\ref{2-1})  as a
cross-diffusion system, in which the second equation contains the
time variable $t$ as a parameter.  The motivation follows from the
well-known fact that application of  Lie's algorithm in the case of
high-order equations leads to very cumbersome formulae. Moreover,
each Lie symmetry of Eq. (\ref{1-1}) can be easily established from
one of system (\ref{2-1})  and there is one-to-one correspondence
between solutions of (\ref{2-1}) and (\ref{1-1}). Thus, we shall
investigate system (\ref{2-1}) instead of Eq. (\ref{1-1}).

Lie's classical algorithm has been implemented within several computer-algebra routines. The
classification problem may be solved by iterating a symmetry-finding program such as the REDUCE-based
program DIMSYM \cite{DimSYM} or MAPLE-based program DESOLV \cite{Vu} that flag special cases of the
coefficient functions that may  give rise to additional symmetries. The REDUCE-based program
 CRACK \cite{CRACK} can fully solve some classification problems automatically.
 {However, these programs may produce many special cases of equations
with additional symmetries, which are equivalent up to the correctly
specified  local substitutions.  Finding and systematically using
the sets of such substitutions often leads to a significant
practical reduction in the number of special cases that admit
additional invariance.
  Thus, the  problem under
 investigation here will be tractable  without the assistance of a
 computer.}

%%%%%%%%%%%%%%%%%%%%%%%%%%%%%%%%%%%%%%%%%%%%%%%%%%%%%%%%%%%%%%%%%%%%
%%%%%%%%%%%%%%%%%%%%%%%%%%%%%%%%%%%%%%%%%%%%%%%%%%%%%%%%%%%%%%%%%%%%%

Now we formulate the main theorem, which presents
 the classification of special forms of system (\ref{2-1}) (with K
not identically 0) having additional symmetries.

\bteo\label{theo}  %@@
All possible  maximal algebras of invariance (up to equivalent
representations generated by transformations of the form
(\ref{2-13})) of  system (\ref{2-1}) for any fixed triplet $(K, \,
D, \, F)$
are presented in  Table 1. Any  other system of the form  %@@
(\ref{2-1}) with non-trivial Lie symmetry
 is reduced
by a local substitution  of the form
\be\label{2-13}  %@@
 \ba  {l} \medskip
 t \to C_0t+C_1e^{C_2t},  \\
\medskip  x \to C_3x, \\
 u \to C_4+  C_5t+ C_6e^{C_7t}u,\\
 \medskip
 v \to C_{8}+  C_{9}e^{C_{10}t}v \ea
    \ee
 to one of those given in Table 1 (the constants  $C$ with
subscripts are determined by the form of the system in question,
some of them necessarily being zero in all particular cases).
\et  %@@

\noindent \textbf{Remark 1}. All the systems listed in Table 1 are
inequivalent up to arbitrary local substitution (not only of the
form (\ref{2-13})!). This can be  shown using the same approach as that used in \cite{ch-se-06} for the general reaction-diffusion-convection equation.

\textbf{Proof.}

 According to the classical Lie scheme \cite{bl-co}--\cite{fss},
we consider system (\ref{2-1}) as the manifold $(S_1,S_2)$ determined by the restrictions:
\be\label{9}
 \ba  {l}
\medskip
S_1\equiv -u_t-[K(u)v_x]_x+[D(u)u_x]_x+F(u)=0,\\
S_2\equiv u_{xx}-v=0 \ea \ee in the space of the variables
 $t, x, u,
v, u_t, v_t, u_x, v_x, u_{xx}, v_{xx}$.
  The maximal algebra of invariance (MAI)
of this system is generated by   infinitesimal  operators of the form
\be\label{2-2}
X=\xi^0(t,x,u,v)\partial_t+\xi^1(t,x,u,v)\partial_x+\eta^1(t,x,u,v)\partial_u+\eta^2(t,x,u,v)\partial_v,
\ee where the functions  $\xi^0, \, \xi^1, \, \eta^1, \, \eta^2$ are
to be determined. In order to determine these  unknown functions one
needs to use the invariance conditions
\be\label{11}  %@@
\ba{l}  %@@
X^{2} S_1  %@@
 \equiv   %@@
X^{2}( -u_t-[K(u)v_x]_x+[D(u)u_x]_x+F(u) )  %@@
\Big\vert_{S_1=0, \, S_2=0}=0, \\[0.3cm]  %@@
X^{2} S_2  %@@
 \equiv  X^{2}  %@@
( u_{xx}-v )  %@@
\Big\vert_{S_1=0, \, S_2=0}=0,  %@@
%{S_1=0}\atop{S_2=0}
\ea  %@@
\ee  %@@
 where $X^{2}$ is  the second  prolongation of the operator $X$:
\be\label{2-3}
 \ba  {l}
\medskip
X^{2}=\xi^0(t,x,u,v)\partial_t+\xi^1(t,x,u,v)\partial_x+\eta^1(t,x,u,v)\partial_u+
\eta^2(t,x,u,v)\partial_v+\\\medskip \quad
+\tau^1_{10}\partial_{u_t}+\tau^1_{01}\partial_{u_x}+\tau^2_{10}\partial_{v_t}+\tau^2_{01}\partial_{v_x}+\\
\quad +\tau^1_{20}\partial_{u_{tt}}+\tau^1_{11}\partial_{u_{tx}}+
\tau^1_{02}\partial_{u_{xx}}+\tau^2_{20}\partial_{v_{tt}}+
\tau^2_{11}\partial_{v_{xt}}+ \tau^2_{02}\partial_{v_{xx}}. \ea \ee
\newpage

{\bf Table 1. Lie symmetries of (\ref{2-1})}
\begin{center}
\begin{tabular}{|c|c|c|c|}
\hline
Case & Systems of the form (\ref{2-1}) & Restrictions & Basic operators of MAI \\
\hline
 &&& \\
 1.& $u_t = -[K(u)v_x]_x$  &  & $P_t, \, P_x$, \\
 & $u_{xx}-v=0$ &  & $D_1=4t\pt_t+x\pt_x-2v\pt_v$\\
\hline
 &&& \\
 2.&  $u_t = -[e^{\gamma u}v_x]_x+d[e^{\mu u}u_x]_x+\lambda e^{(2\mu-\gamma)u}$
  & $\gamma^2+\mu^2\neq0$ &  $P_t, \, P_x$, \\
& $u_{xx}-v=0$ & $d^2+\la^2\neq0$ & $ D_2=2(\gamma-2\mu)t\pt_t+(\gamma-\mu)x\pt_x$\\
& & & $+2(\pt_u-(\gamma-\mu)v\pt_v)$ \\
\hline
& & & \\
3.&  $u_t = -[u^\gamma v_x]_x+d[u^\mu u_x]_x+\lambda u^{2\mu-\gamma+1}$  & $\gamma^2+\mu^2\neq0$  &  $P_t, \, P_x$, \\
 & $u_{xx}-v=0$ & $d^2+\la^2\neq0$ & $ D_3=2(\gamma-2\mu)t\pt_t+(\gamma-\mu)x\pt_x$\\
& & &  $+2(u\pt_u+(\mu-\gamma+1)v\pt_v)$ \\
 \hline
& & & \\
4.&  $u_t = -v_{xx}+du_{xx}+\lambda u\ln u$ &  $\lambda\neq0$ & $P_t, \, P_x$, \\
 & $u_{xx}-v=0$ &  & $Q_1=e^{\lambda t}(u\pt_u+v\pt_v)$  \\
\hline
&&& \\
5.& $u_t = -[e^{\gamma u}v_x]_x$  & $\gamma\neq0$ & $P_t, \, P_x$, \\
 & $u_{xx}-v=0$ &  & $D_1,\quad D_2$ with $\mu=\gamma$\\
\hline
&&& \\
6.& $u_t = -[u^\gamma v_x]_x$  &  $\gamma\neq0$ & $P_t, \, P_x$, \\
 & $u_{xx}-v=0$ &  & $D_1, \quad D_3$ with  $\mu=\gamma$\\
\hline
&&& \\
7.& $u_t = -v_{xx}+du_{xx}$  & $d\neq0$ & $P_t, \, P_x, \, I=u\pt_u+v\pt_v$, \\
 & $u_{xx}-v=0$ &  & $X^\infty=P(t,x)\pt_u+P_{xx}(t,x)\pt_v$, \\
& & & where $P_t+P_{xxxx}-dP_{xx}=0$\\
\hline
&&& \\
8. &  $ u_t = -v_{xx}  $ &   & $P_t, \, P_x, \, I, \, D_1, $ \\
  &$u_{xx}-v=0$   &  &
$X^\infty=P(t,x)\pt_u+P_{xx}(t,x)\pt_v$, \\
& & & where $P_t+P_{xxxx}=0$ \\
\hline
\end{tabular}
\end{center}

\vspace{2cm}

The coefficients with relevant subscripts
    $\tau^1_t$, $\tau^1_x$ ,
$\tau^2_{10}$, $\tau^2_{01}$, $\tau^1_{20}$, $\tau^1_{11}$,
$\tau^1_{02}$, $\tau^2_{20}$, $\tau^2_{11}$, $\tau^2_{02}$ are
calculated by the well-known prolongation formulae (see, e.g.
\cite{bl-co}--\cite{fss}).

Substituting  operator (\ref{2-3}) into system (\ref{11}) and
carrying out  the relevant
%cumbersome
 calculations, we obtain
 so called system of
determining equations for finding  the functions  $\xi^0, \, \xi^1,
\, \eta^1, \, \eta^2$. This is an overdetermined  system of PDEs
that can be written in the explicit form:
 \be\label{13}
\xi^0_x=\xi^0_u=\xi^0_v=\xi^1_t=\xi^1_u=\xi^1_v=\xi^1_{xx}=0\\
\ee \be\label{14}
\eta^1_v=\eta^1_{xu}=\eta^1_{uu}=0\\
\ee \be\label{15} \eta^2=\eta^1_{xx}+(\eta^1_u-2\xi^1_x)v\\ \ee
%%\be\label{2-7} K'(u)\eta^1_x=0\\ \ee
 \be\label{16}
K(u)(\xi^0_t-4\xi^1_x)+K'(u)\eta^1=0\\
\ee \be\label{17} D(u)(\xi^0_t-2\xi^1_x)+D'(u)\eta^1=0 \ee
\be\label{18}
\eta^1_t+K(u)\eta^1_{xxxx}+F(u)(\eta^1_u-\xi^0_t)-D(u)\eta^1_{xx}-F'(u)\eta^1=0\\
\ee

Of course,  Eqs.  (\ref{13})--(\ref{15}) are linear and don't depend
on the functions $K, \, D$ and $F$, hence their general solution can
be easily constructed: \be \label{19}
 \ba  {l}
\medskip
\xi^0=\xi^0(t),\\
\xi^1=\al x+x_0,\\
\eta^1=R(t)u+P(t,x),\\
\eta^2=P_{xx}(t,x)+(R(t)-2\al)v, \ea \ee where $P(t,x)$, $R(t)$  are
arbitrary smooth functions, $\al$ and  $x_0$  are arbitrary
constants.
 Eqs.
(\ref{16})--(\ref{18})  form the system of classification equations.
Its general  solution under assumption  of arbitrarily given
functions $K, \, D$ and $F$ generates the invariance algebra that
consists only of generators of translations in $x$ and $t$ with the
basis
%\be\label{2-11} \xi^0= e_1, \quad \xi^1=e_2, \quad
%\eta^1=\eta^2=0. \ee
\be\label{2-12} P_t = \frac{\ts\pt}{\ts \pt t}\equiv \pt_t,
 \, P_x =\frac{\ts\pt}{\ts \pt x}\equiv \pt_x.
 \ee%(4)
The algebra with this basis  is called  {\it the trivial  %@@
Lie algebra} of the system (\ref{2-1}) (note that other authors,
instead use `kernel of the basic Lie  groups'\cite{ovs} or 'the
principal Lie algebra'\cite{gan-ibr-2008} in this context).
Thus, we aim to find all  triplets $(K, \, D, \, F)$  %@@
 that lead to extensions of
 the trivial Lie algebra generated by (\ref{2-12}). This means that
 one needs to solve Eqs.
(\ref{16})--(\ref{18}) with coefficients (\ref{19}). The crucial
step in solving this task    is to analyze
 differential consequences of Eqs.
(\ref{16})--(\ref{17}) with respect to the variable $x$.
 Since these consequences take the form
 $K'(u)P_x(t,x)=0$ and  $
D'(u)P_x(t,x)=0$, respectively,  one arrives at two basic cases \[
\textbf{i} \ [K'(u)]^2+[D'(u)]^2\neq 0 \qquad \qquad  \textbf{ii} \
K'(u)=D'(u)=0 .\]

Consider case \textbf{i}. In this case   $P_x(t,x)=0$, so that Eqs.
(\ref{16})--(\ref{18}) take the form  \be\label{23}
K(u)[\xi^0_t(t)-4\al]+K'(u)[R(t)u+P(t)]=0, \ee \be\label{24}
D(u)[\xi^0_t(t)-2\al]+D'(u)[R(t)u+P(t)]=0,
\ee
\be\label{25} R_t(t)u+P_t(t)+F(u)[R(t)-\xi^0_t(t)]-F'(u)[R(t)u+P(t)]=0.\\
\ee

Setting  $R(t)=P(t)=0$, we immediately arrive at case 1 of Table 1.
In fact, Eqs. (\ref{23})-(\ref{25}) with $R(t)=P(t)=0$ and non-zero
$K(u)$ are equivalent to
\[ \xi^0_t(t)=4\al \not=0, \quad  D(u)=0, \quad F(u)=0\]
and $K(u)$ is an arbitrary smooth function. This means that the
triplet $(K(u), 0, 0)$ forms the system from case 1 of Table 1 and
the coordinates of the infinitesimal operator (\ref{2-2}) take the
form \be \label{19*}
 \ba  {l}
\medskip
\xi^0=4\al t +t_0, \quad
\xi^1=\al x+x_0,\\
\eta^1=\eta^2=0, \ea \ee where $\al, t_0$ and $x_0$ are  arbitrary
parameters. Operator (\ref{2-2}) with coordinates (\ref{19*})
generates exactly the operators  $P_t$ (for $ t_0=1, \al=x_0=0$),
$P_x$ (for $ t_0= \al=0, x_0=1$) and $D_1$ (for $\al=1, x_0=t_0=0$)
listed in case 1 of Table 1.

 If $R^2(t)+P^2(t) \neq0$ then two possible subcases arise:
\textbf{ia.} $R(t) \neq0$ and  \textbf{ib.} $R(t)=0, \ P(t) \neq0$.

Consider subcase  \textbf{ia}. Integrating Eq. (\ref{23}) as an ODE
on the function $ K(u)$, one obtains \be\label{127} K(u)=k
\Bigl[u+\frac{P(t)}{R(t)}\Bigr]^{\frac{4\al-\xi^0_t(t)}{R(t)}}, \ee
where $k$ is a nonzero constant, which can be reduced to  $k=1$ (by
scaling time $t \to kt$) without losing generality.
 Since the function $K$ must depend only on
$u$, the restrictions \be \label{132} \ba{l}
\medskip
\xi^0_t(t)=4\al-\ga R(t),\\
P(t)=\ga_0R(t) \ea \ee are obtained. Here  $\ga$ and $ \ga_0$ are
arbitrary constants. Thus, solving Eq. (\ref{23}), we arrive at the
power function \be\label{127*} K(u)=(u+\ga_0)^\ga \ee and
restrictions (\ref{132}).

To solve Eqs. (\ref{24})-(\ref{25}), we need to analyze two subcases
\textbf{ia1} $D(u)\not=0$ and  \textbf{ia2} $D(u)=0$.

%the differential consequence of Eq. (\ref{24}) with respect to the
%variable  $t$. Since this is  $R_t(t)\al D(u)=0$ we should analyze
%three possibilities: \textbf{ia1.} $R_t(t)=0$, \qquad \textbf{ia2.}
%$R_t(t) \neq0 \Rightarrow \al=0$ \qquad and \qquad \textbf{ia3.}

In  subcase  \textbf{ia1},  Eq. (\ref{24}) can be solved as an  ODE
on the function $ D(u)$, so that one obtains \be\label{128} D(u)= d
(u+\ga_0)^\mu \ee and the condition \be \label{129} \
\xi^0_t(t)=2\al-\mu R(t), \ee where $d \not=0$ and $\mu$ are
arbitrary constants.

If $\ga \not=\mu$ then solving Eqs.  (\ref{132}) and (\ref{129}) and
substituting the found functions $R(t),P(t)$ and $\xi^0$ into Eq.
(\ref{25}), we arrive at the ODE \be\label{130}
(u+\ga_0)F'(u)+(\ga-2\mu-1)F(u)=0. \ee The general solution of Eq.
(\ref{130}) and the functions $ K(u)$ and $ D(u)$ found above form
the system \be\label{131} \ba{l} \medskip u_t = -[(u+\ga_0)^\ga
v_x]_x+
d[(u+\ga_0)^\mu u_x]_x+\la (u+\ga_0)^{2\mu-\ga +1} \\
 u_{xx}-v=0,  \ea  \ee
 which is reduced to the system listed in case 3 of Table 1 by
 renaming $u+\ga_0 \to u$. Substituting the functions $R(t),P(t)$ and $\xi^0$
 into (\ref{19}), we obtain the infinitesimal  operator (\ref{2-2}),
which generates three basic operators listed in case 3 of Table 1.

If $\ga =\mu$ then Eqs.  (\ref{132}) and (\ref{129}) are compatible
only under restriction $\al=0$. It turns out that this restriction
doesn't lead to any new cases but to case 3 of Table 1 with $\ga
=\mu$. In fact, Eqs. (\ref{23}) and (\ref{24}) have identical
structure, hence
 $D(u)=d(u+\ga_0)^\ga$, $d=const$.  The general solution of
(\ref{25}) has the form  $F(u)=\la(u+\ga_0)^{1+\ga}+\la_1(u+\ga_0)$,
where  $\la$ and  $\la_1\neq 0$ are arbitrary constants. The relevant
coordinates of infinitesimal  operator (\ref{2-2}) take the form \be
\label{133} \ba{l} \medskip \xi^0=\frac{r_1}{\la_1}e^{-\la_1 \ga
t}+t_0, \,
 \xi^1=x_0,\\
\eta^1=r_1e^{-\la_1 \ga t}(u+\ga_0), \, \eta^2=r_1e^{-\la_1 \ga t}v,
\ea\ee where $t_0, r_1$ and $x_0$ are arbitrary parameters. Thus,
the system \be\label{134} \ba{l} \medskip u_t = -[(u+\ga_0)^\ga
v_x]_x+
d[(u+\ga_0)^\ga u_x]_x+\la (u+\ga_0)^{1+\ga}+\la_1 (u+\ga_0) \\
 u_{xx}-v=0  \ea  \ee
is invariant under three-dimensional MAI with the basic operators
\be\label{135} \ba{l} P_t, P_x, \\
 Q^*=e^{-\gamma \la_1 t}(\pt_t+\la_1((u+\ga_0)\pt_u+v\pt_v)). \ea  \ee
However, we found the local substitution
\be\label{136} \ba{l} t^*=\frac{1}{\la_1\ga}e^{\la_1 \ga t}, \\
x^*=x,\\
u^*=(u+\ga_0)e^{-\la_1 t},\\
v^*=ve^{-\la_1 t}, \ea \ee which reduces system (\ref{134}) and Lie
algebra (\ref{135}) to  the system and Lie algebra listed in case 3
of Table 1 with $\ga =\mu$.

 The  analysis of subcase \textbf{ia2} is straightforward because
 Eq. (\ref{24}) vanishes for $D(u)=0$ while Eq. (\ref{25})
 can be treated in a similar way. In conclusion,
 we found that  the system
\be\label{137} \ba{l} \medskip u_t = -[(u+\ga_0)^\ga v_x]_x+\la_1 (u+\ga_0), \\
 u_{xx}-v=0,  \ea  \ee
is invariant under four-dimensional MAI with the basic operators
\be\label{138} \ba{l} P_t, P_x, Q^*, \\
D^*=x\pt_x+\frac{4}{\ga}(u+\ga_0)\pt_u+(\frac{4}{\ga}-2)v\pt_v. \ea
\ee Direct checking shows that system (\ref{137}) and Lie algebra
(\ref{138}) are reduced  to  the system and Lie algebra listed in
case 6 of Table 1 if one applies    substitution (\ref{136}).

Examination of  subcase  \textbf{ib}  is much simpler because Eqs.
(\ref{23})--(\ref{25}) with  $R(t)=0$ can be easily integrated.
Finally, one obtains cases 2 and 5 of Table 1.

To complete the proof we need to examine  case \textbf{ii}
$K'(u)=D'(u)=0 $, i.e.  $K(u)=k=const$, $D(u)=d=const$. Since $K(u)
\neq0$, we can again set  $k=1 $  without losing  generality.

%\newpage

Thus, the  classification equations (\ref{16})--(\ref{17}) with
coefficients (\ref{19}) can be essentially simplified and one
obtains
 \be \label{28} \xi^0(t)=4\al t+t_0, \qquad \al d=0. \ee
The third  classification equation (\ref{18}) takes the form
\be\label{29} \ba {l}
\medskip
R_t(t)u+P_t(t,x)+P_{xxxx}(t,x)-dP_{xx}(t,x)+\\
+F(u)[R(t)-4\al]-F'(u)[R(t)u+P(t,x)]=0. \ea \ee Differentiating Eq.
(\ref{29}) with respect to  $x$ and  $u$, we find the condition
$F''(u)P_x(t,x)=0$. Hence two different subcases should be examined:
\[\textbf{iia} \ F'' = 0  \qquad  \textbf{iib} \  F'' \neq 0, \ P_x(t,x)=0.\]
%%  $P(t,x)=P(t)$.

Consider  subcase \textbf{iia}. Since $F''(u)=0$ we immediately
obtain  $F(u)=\la_1u+\la_0$ so that the  triplet  of functions $(K,
\, D, \, F)$  is known.

Substituting the function $F$ into Eq. (\ref{29}) and  splitting the
obtained expression into two equations (with the variable $u$ and
without it) we arrive at  $R(t)=4\al \la_1 t+r_1$ and the linear
PDE \be \label{40}
P_t(t,x)+P_{xxxx}(t,x)-dP_{xx}(t,x)-\la_1P(t,x)+\la_0(4\al \la_1
t+r_1-4\al)=0 \ee to find the function $P(t,x)$. Thus, the
coordinates of infinitesimal  operator (\ref{2-2}) take the form
\be \label{39} \ba{l} \medskip \xi^0=4\al t+t_0, \, \xi^1=\al x+x_0,\\
\eta^1=(4\al \la_1 t+r_1)u+P(t,x), \, \eta^2=(4\al \la_1
t+r_1-2\al)v+P_{xx}(t,x), \ea\ee where  $P(t,x)$ is the general
solution of  Eq. (\ref{40}).

The last step is to take into account the second condition from
(\ref{28}). If $d\neq 0$ then  $\al=0$ and, applying the relevant
simplifications, we arrive at the case 7 of Table 1.
% Note Eq. (\ref{40}) with $\al=0$ is reduced to one
%presented in case 7 of table 1.

 If  $d=0$ then  Eq. (\ref{40}) takes the
form \be \label{41} P_t(t,x)+P_{xxxx}(t,x)-\la_1P(t,x)+\la_0(4\al
\la_1 t+r_1-4\al). \ee The corresponding system is
 \be \label{45}
\ba{l}
\medskip
u_t=-v_{xx}+\la_1u+\la_0,\\
u_{xx}-v=0. \ea \ee
 It turns out that  system
(\ref{41}) is reduced to the form  listed in  case 8 of Table 1 if
one applies the local substitutions
 \be \label{42} \ba{l}
\medskip
u^*=u-\la_0 t, \quad \la_1=0  \\
v^*=v \ea \ee and
  \be \label{44} \ba{l}
\medskip
u^*=e^{-\la_1t}(u+\frac{\la_0}{\la_1}), \quad \la_1\not=0 \\
v^*=e^{-\la_1 t}v. \ea \ee Simultaneously, operator (\ref{2-2}) with
coordinates (\ref{39}) is transformed in such a way that the basic
operators $P_t, \, P_x, \, I, \, D_1 $ and
$X^\infty=P(t,x)\pt_u+P_{xx}(t,x)\pt_v$, listed in case 8 of Table
1, can be easily derived.

Consider  subcase \textbf{iib}. Since $P_x(t,x)=0$, Eq. (\ref{29})
takes the form \be\label{30}
R_t(t)u+P_t(t)+F(u)[R(t)-4\al]-F'(u)[R(t)u+P(t)]=0.\\
\ee Differentiating Eq. (\ref{30}) with respect to the variables $t$
and $u$, we find the equation
\be\label{30*}F''(u)[R_t(t)u+P_t(t)]=R_{tt}(t).\ee
 Because Eq.
(\ref{30*}) has a simple structure we prefer to solve this equation
and check when the solution obtained will satisfy Eq. (\ref{30}). We
note that the special case  $R_t(t)=P_t(t)=0$ doesn't lead to new
results, so that  $R_t^2(t)+P_t^2(t) \neq0$.
 % \be \label{31}
% F''(u)=\frac{R_{tt}(t)}{R_t(t)u+P_t(t)}. \ee
Moreover, since the function $F$ depends only on $u$, the relations
\be \label{32} \ba{l} \medskip
P_t(t)=\ga R_t(t),\\
R_{tt}(t)=\la R_t(t), \ea \ee where  $\ga$ and $\la\neq 0$ are
arbitrary constants, should take place. The general solution of Eq.
(\ref{30*}) with the coefficients  (\ref{32}) is
 \be \label{34}
F(u)=\la(u+\ga)\ln(u+\ga)+\la_1 u+\la_0, \ee where  $\la_0$ and
$\la_1$ are arbitrary constants. Now we substitute (\ref{34}) and
the general solution of the linear ODEs system (\ref{32}) into Eq.
(\ref{30}) and find conditions when the obtained expression can be
fulfilled. The simple calculations give
 \be \label{35} \ba{l}
\medskip R(t)=r_1e^{\la t}, \qquad
P(t)=\ga r_1e^{\la t} \\
\al=0, \qquad  \la_0=\la_1 \ga. \ea \ee

So  the system
 \be \label{36}
\ba{l}
\medskip
u_t=-v_{xx}+du_{xx}+\la(u+\ga)\ln(u+\ga)+\la_1(u+\ga),\\
u_{xx}-v=0 \ea \ee admits MAI generated by operator (\ref{2-2}) with
coordinates
\be \label{37} \ba{l} \medskip \xi^0=t_0, \, \xi^1=x_0,\\
\eta^1=r_1e^{\la t}(u+\ga), \, \eta^2=r_1e^{\la t}v. \ea\ee
Finally,
the system  (\ref{36}) and operator (\ref{2-2}) with (\ref{37}) are
simplified by the substitution
 \be \label{38} \ba{l}
\medskip
u^*=e^{\frac{\la_1}{\la}}(u+\ga),\\
v^*=e^{\frac{\la_1}{\la}}v,
 \ea \ee
so that the system  and MAI listed in case 4 of Table 1 are
obtained.

Thus, the system of determining equations (\ref{13})---(\ref{18}) is
completely  solved and eight different systems  of the form
(\ref{2-1}) have been  found, which admit three- and
higher-dimensional  Lie algebras. Simultaneously, we have shown that
all other systems admitting non-trivial Lie algebra are reduced to
those listed in Table 1  by the substitutions of the form
 (\ref{136}), (\ref{42}), (\ref{44}) and
(\ref{38}). One notes that all of these substitutions can be united to
the form (\ref{2-13}).

The proof is now completed. \hfill $\blacksquare$

%%%%%%%%%%%%%%%%%%%%%%%%%%%%%%%%%%%%%%%%%%%%%%%%%%%%%%%%%
\bigskip

One easily notes that cases 2 and 3 of  Table 1 generalize the
results of Lie symmetry analysis for the Cahn-Hilliard equation
derived in \cite{choudhury-95,gandar-2000}. For example, the systems
\be\label{2-14} \ba{l} \medskip
u_t = -v_{xx}+d[e^{\mu u}u_x]_x, \\
u_{xx}-v=0
\ea \ee
and
\be\label{2-15} \ba{l} \medskip
u_t = -v_{xx}+d[u^\mu u_x]_x, \\
u_{xx}-v=0, \ea \ee which are the particular cases of the
corresponding systems from Table 1,  admit the Lie symmetry
operators \be\label{2-16} 4\mu t\pt_t+\mu x\pt_x-2(\pt_u+\mu
v\pt_v)\ee and \be\label{2-17} 4\mu t\pt_t+\mu x\pt_x
-2(u\pt_u+(\mu+1)v\pt_v) \ee respectively. The table also includes the results obtained in \cite{gandar-2000} for
the equation Eq.  (\ref{1-1}) with  $K(u)=const$ and  $F(u)=0$.

 To finish the Lie symmetry description we  note that Eq. (\ref{1-1}) can be reduced
 to an equivalent system of PDEs in different ways.
 One sees that the system
\be\label{2-1*} \ba{l} \medskip
u_t = -[K(u)v_x]_x+[D(u)u_x]_x+F(u),\\
0= u_{x}-w,\\
 0= w_{x}-v \ea \ee
  by introducing new unknown functions  $v=v(t,x), w=w(t,x)$ can be obtained from Eq. (\ref{1-1}). Since the system includes the first-order variable $w=u_x$, point symmetries of this system include contact symmetries of the original single fourth-order  equation.
System (\ref{2-1*})  is nothing else but a cross-diffusion system,
in which the second and third  equations contain the time variable
$t$ as a parameter. Thus, we may investigate also system
(\ref{2-1*}) instead of Eq. (\ref{1-1}). There is an essential
difference between systems (\ref{2-1}) and (\ref{2-1*}) because the
second system contains the equations of different orders.  In fact,
 according to the classical Lie scheme,  MAI
of this system is generated by the infinitesimal operator
\be\label{2-2*} \ba  {l}
X=\xi^0(t,x,u,v,w)\partial_t+\xi^1(t,x,u,v,w)\partial_x+\eta^1(t,x,u,v,w)\partial_u+
\\
+\eta^2(t,x,u,v,w)\partial_v++\eta^3(t,x,u,v,w)\partial_w, \ea \ee
where the functions  $\xi^0, \, \xi^1, \, \eta^1, \, \eta^2, \eta^3$
are to be determined. Applying the second
 prolongation of the operator  (\ref{2-2*}) to system (\ref{2-1*})
 and using the invariance
conditions   one can derive the determining equations for finding
the functions  $\xi^0, \, \xi^1, \, \eta^1, \, \eta^2, \eta^3$. It
should be stressed that the relevant invariance conditions must take
into account the differential consequences (with respect to the
variable $x$) of the second and third equations of system
(\ref{2-1*}). We omit cumbersome calculations and present the final
result in the explicit form:

\be\label{2-4*}
\xi^0_x=\xi^0_u=\xi^0_v=\xi^1_t=\xi^1_u=\xi^1_v=\xi^1_{xx}=0\\
\ee \be\label{2-5*}
\eta^1_v=\eta^1_w=\eta^1_{xu}=\eta^1_{uu}=0\\
\ee \be\label{2-6*}
\eta^2=\eta^1_{xx}+(\eta^1_u-2\xi^1_x)v\\
\ee \be\label{2-6**}
\eta^3=\eta^1_{x}+(\eta^1_u- \xi^1_x)w\\
\ee \be\label{2-8*}
K(u)(\xi^0_t-4\xi^1_x)+K'(u)\eta^1=0\\
\ee \be\label{2-9*} D(u)(\xi^0_t-2\xi^1_x)+D'(u)\eta^1=0 \ee
\be\label{2-10*}
\eta^1_t+K(u)\eta^1_{xxxx}+F(u)(\eta^1_u-\xi^0_t)-D(u)\eta^1_{xx}-F'(u)\eta^1=0\\
\ee

Now we note that the determining equations obtained (of course,
without Eq. (\ref{2-6**})) are equivalent to those for the system
(\ref{2-1}) so that no new Lie point symmetries or contact symmetries can be found.

\begin{center}
{\textbf{3. Symmetry reduction and  exact solutions }}
\end{center}

Some of the systems presented in Table 1  are equivalent to  known
fourth order PDEs arising in applications.  For example, the system
listed in case 6 is nothing else but the thin film equation
(\ref{1-3}). Since the motivation to this study is to consider this
type of equation  with non-zero reaction terms, which naturally
arise in some processes \cite{segel-84}-\cite{Gaskell}, henceforth
we restrict our attention mainly to case 3 of Table 1.

It is well-known that a Lie symmetry allows one to reduce the given
PDE (system of PDEs) to an equation (system of equations) of lower
dimensionality.  Here we reduce systems arising in  case 3 of
Table 1 to systems of ordinary differential equations (ODEs),
furthermore these ODE systems are solved in particular cases and
exact solutions of the initial PDE systems are constructed. Finally,
these solutions  are compared with those obtained by other authors.

Consider the system arising in the case 3 of Table 1:
\be\label{3-1} \ba{l} \medskip
u_t = -[u^\ga v_x]_x+d[u^\mu u_x]_x+\la u^{2\mu-\ga+1}, \\
%%=- u^{\ga-1}[\ga u_xv_x+uv_{xx}]+du^{\mu-1}[\mu u_x^2+uu_{xx}]+ \la u^{2\mu-\ga+1}$, \quad
 u_{xx}-v=0,\ea \ee
 where $\ga^2+\mu^2\neq 0$.
 The most general Lie symmetry operator of  system (\ref{3-1})
has the form
\be\label{3-2} \ba{l} \medskip
X=\al_1P_t+\al_2P_x+\al_3D_3=\\
=[\al_1+2\al_3(\ga-2\mu)t]\pt_t+[\al_2+\al_3(\ga-\mu)x]\pt_x+[2\al_3u]\pt_u+
[2\al_3(1-\ga+\mu)v]\pt_v, \ea \ee where $\al_i, i=1,2,3$ are
arbitrary constants. To construct the relevant ansatz one needs to
solve the Pfaffian system of characteristic equations
 \be\label{3-3}
\frac{dt}{\al_1+2\al_3(\ga-2\mu)t}=\frac{dx}{\al_2+\al_3(\ga-\mu)x}=\frac{du}{2\al_3u}=\frac{dv}{2\al_3(1-\ga+\mu)v}.\ee
The general solution of (\ref{3-3}) essentially depends on the
parameters $\al_1, \al_2, \al_3, \ga$ and $\mu$ and five different
cases occur.

 \textbf{Case 1} $\al_3=0$ leads to the plane wave solutions of the form
\be\label{3-4}  \om=\al_1x-\al_2t, \quad u=\phi(\om),\quad
v=\psi(\om),\ee where $\phi$ and $\psi$ are new unknown functions.
These functions should satisfy the ODE system \be\label{3-5} \ba{l}
\medskip
-\al_2\phi'=-\al_1^2\phi^{\ga-1}[\ga\phi'\psi'+\phi\psi'']
+d\al_1^2\phi^{\mu-1}[\mu(\phi')^2+\phi\phi'']+\la\phi^{1-\ga+2\mu},\\
\psi=\al_1^2\phi''.
\ea \ee \\
It should be noted that this system is equivalent to the fourth-order ODE
\be\label{3-6} \ba{l}
\medskip
\al_1^4\phi^{\ga}\phi^{iv}+\ga\al_1^4\phi^{\ga-1}\phi'\phi'''-
d\al_1^2\phi^\mu\phi''-d\al_1^2\mu\phi^{\mu-1}(\om)(\phi')^2-
\al_2\phi'-\la\phi^{1-\ga+2\mu}=0. \ea \ee This equation is not
integrable because there are no general solutions for this equation
in terms of elementary functions and known special functions
\cite{pol-za}.  However, it can be noted that the special case with
$\ga=3\mu$ possesses the particular solution
 \be\label{51*}
\phi(\om)=\al\om^{\frac{1}{\mu}}, \ee where
 $\al$ is a solution of algebraic equation
\be\label{51**}
\al_1^4(1-\mu)(1-2\mu)\al^{4\mu}-d\al_1^2\mu^2\al^{2\mu}-\al_2\mu^3\al^\mu-\la
\mu^4=0. \ee  Hence the system  \be\label{51***} \ba{l} \medskip
u_t = -[u^{3\mu} v_x]_x+d[u^\mu u_x]_x+\la u^{1-\mu}, \\
u_{xx}-v=0 \ea \ee possesses the exact  solution
 \be\label{51****} \ba{l}
\medskip
u=\al(\al_1x-\al_2t)^\frac{1}{\mu},\\
v=\al_1^2\al\frac{1}{\mu}(\frac{1}{\mu}-1)(\al_1x-\al_2t)^{\frac{1}{\mu}-2},\\
\ea \ee  where   $\al$ satisfies   (\ref{51**}).

\textbf{Case 2} $\al_3\neq0$, \quad $\ga=2\mu\neq0$,  \quad
$\al_1=0$ \ leads to the ansatz \be\label{3-7}  \om=t, \quad
u=\phi(t)x^{\frac{2}{\mu}},\quad v=\psi(t)x^{\frac{2}{\mu}-2}. \ee
The corresponding ODE system takes the form \be\label{3-8} \ba{l}
\medskip
\phi'=-2(\frac{1}{\mu}-1)(1+\frac{2}{\mu})\phi^{2\mu}\psi+
d\frac{2}{\mu}(\frac{2}{\mu}+1)\phi^{\mu+1}+\la\phi,\\
\psi=\frac{2}{\mu}(\frac{2}{\mu}-1)\phi \ea \ee and can be rewritten
as the single ODE: \be\label{3-9}
\phi'=-\frac{4}{\mu}(\frac{1}{\mu}-1)(\frac{4}{\mu^2}-1)\phi^{2\mu+1}+
d\frac{2}{\mu}(\frac{2}{\mu}+1)\phi^{\mu+1}+\la\phi. \ee

It should be noted that  this   ODE with $\lambda=0$ coincides with
one derived in the recently published paper \cite{gan-ibr-2008} for
the fourth-order PDE, which is equivalent to  (\ref{3-1}) with
$\lambda=0$. However, system (\ref{3-1}) with $\lambda \not=0$
cannot be reduced to one with $\lambda=0$ so that the solutions
presented below  cannot be obtained from \cite{gan-ibr-2008}.

 If $\mu=1$
or $\mu=2$ then  the known solutions of the reaction-diffusion
equations
\[u_t=d[uu_x]_x+\la u \qquad \mu=1,\]
and
\[u_t=d[u^2u_x]_x+\la u  \qquad \mu=2\]
are obtained because $u_{xxx}=0$ (see (\ref{3-7})).
If  $\mu=-2$ then
\be\label{3-10} \phi(t)=Ce^{\la t} \ee
and there follows an exact solution
\be\label{3-12} \ba{l}
\medskip
u=\frac{Ce^{\la t}}{x}, \\
v=\frac{2Ce^{\la t}}{x^3} \ea \ee
of the system
\be\label{3-11} \ba{l} u_t = -[\frac{1}{u^4} v_x]_x+d[\frac{1}{u^2} u_x]_x+\la u, \\
u_{xx}-v=0. \ea \ee

If  $\mu\neq 1; \pm2 $ then two subcases, $\la=0$ and $\la\neq0$,
should be separately considered. Both of them lead to the function
$\phi(t)$ in the implicit form and one can be found from the
transcendental equation \be\label{3-13}
\ln\Bigl|1+\frac{d}{2(1-\frac{1}{\mu})(\frac{2}{\mu}-1)}\frac{1}{\phi^\mu}\Bigr|-
\frac{d}{2(1-\frac{1}{\mu})(\frac{2}{\mu}-1)}\frac{1}{\phi^\mu}=
\frac{d^2(\frac{2}{\mu}+1)}{(1-\frac{1}{\mu})(\frac{2}{\mu}-1)}(t-t_0)\\
 \ee
 if $\la=0$ and from the equation
 \be\label{3-13*}
\ln\frac{\phi^{2\mu}(t)}{|\phi^{2\mu}(t)+\al\phi^\mu(t)+\beta|}-\al\int
\limits_{0}^{\phi^\mu(t)} \frac{dz}{z^2+\al z+\beta}=2\la\mu
(t-t_0),\
 \ee

$\al=\frac{d\mu^2}{2(1-\mu)(\mu-2)}$,
$\beta=\frac{\la\mu^4}{4(1-\mu)(\mu^2-4)}$\\
if $\la \not=0$.\\

 Thus, the system
\be\label{3-14} \ba{l} \medskip
u_t = -[u^{2\mu} v_x]_x+d[u^\mu u_x]_x, \\
u_{xx}-v=0 \ea \ee
possesses the exact solution
\be\label{3-15} \ba{l}
\medskip
u=\phi(t)x^{\frac{2}{\mu}},\\
v=\frac{2}{\mu}(\frac{2}{\mu}-1)\phi(t)x^{\frac{2}{\mu}-2}, \ea
\ee where  $\phi(t)$ satisfies the equation (\ref{3-13}). Note
that the function $\phi(t)$ tends to $0$ if $t \to \infty$ and
this function blows up if $t \to t_0$. These properties  follow
from the simple
 analysis of  (\ref{3-13}).\\
Analogously, the system \be\label{3-14-} \ba{l} \medskip
u_t = -[u^{2\mu} v_x]_x+d[u^\mu u_x]_x+\la u, \\
u_{xx}-v=0 \ea \ee possesses the exact solution (\ref{3-15}) with
$\phi(t)$ satisfying the equation
(\ref{3-13*}).\\

\textbf{Case 3} $\al_3\neq0$, \quad $\ga=2\mu\neq0$,  \quad
$\al_1\neq0$ leads to the ansatz \be\label{3-16} \qquad
\om=xe^{-\frac{\al_3\mu}{\al_1}t}, \quad
u=\phi(\om)e^{\frac{2\al_3}{\al_1}t}, \quad
v=\psi(\om)e^{\frac{2\al_3}{\al_1}(1-\mu)t}, \ee which reduces the
initial system to the ODE system \be\label{3-17} \ba{l}
\medskip
-\al_3\mu\om\phi'+2\al_3\phi=-\al_1\phi^{2\mu-1}[2\mu\phi'\psi'+\phi\psi'']+
d\al_1\phi^{\mu-1}[\mu(\phi')^2+\phi\phi'']+\al_1\la\phi,\\
\psi=\phi''.
\ea \ee \\

System (\ref{3-17}) is equivalent to the 4-th order equation
\be\label{3-18} \ba{l}
\medskip
\al_1\phi^{2\mu}\phi^{iv}+2\mu\al_1\phi^{2\mu-1}\phi'\phi'''-
d\al_1\phi^\mu\phi''-d\al_1\mu\phi^{\mu-1}(\om)(\phi')^2-\\
-\al_3\mu\om\phi'-(\al_1\la-2\al_3)\phi=0, \ea \ee which
 possesses
the particular solution \be\label{3-18**}
\phi(\om)=\al\om^{\frac{2}{\mu}}. \ee  Here,
 $\al$ must be a solution of the algebraic equation
\be\label{3-18***}
4(4-\mu^2)(1-\mu)\al^{2\mu}-2d\mu^2(\mu+2)\al^\mu-\la\mu^4=0, \ee
which is simply a quadratic equation for $\al^{\mu}$.

Thus,  the cross-diffusion system
\be\label{3-14*} \ba{l} \medskip
u_t = -[u^{2\mu} v_x]_x+d[u^\mu u_x]_x+\la u, \\
u_{xx}-v=0 \ea \ee
has the stationary solution
\be\label{3-14**} \ba{l} \medskip
u=\al x^{\frac{2}{\mu}}, \\
v=2\al\frac{2-\mu}{\mu^2}x^{\frac{2}{\mu}-2}, \ea \ee
where
$\al$  satisfies (\ref{3-18***}).

\textbf{Case 4} $\al_3\neq0$,  \quad  $\ga=\mu \neq0$ leads to the
ansatz \be\label{3-19} \om=x+\frac{\al_2}{2\mu\al_3}\ln t, \quad
u=\phi(\om)t^{-\frac{1}{\mu}}, \quad v=\psi(\om)t^{-\frac{1}{\mu}}.
\ee The corresponding ODE system takes the form \be\label{3-20}
\ba{l}
\medskip
\al_2\phi'-2\al_3\phi=-2\al_3\mu\phi^{\mu-1}[\mu\phi'\psi'+\phi\psi'']+
2\al_3\mu
d\phi^{\mu-1}[\mu(\phi')^2+\phi\phi'']+2\al_3\mu\la\phi^{\mu+1},\\
\psi=\phi'' \ea \ee and is equivalent to the 4-th order equation

\be\label{3-21} \ba{l}
\medskip
2\al_3\mu\phi^\mu\phi^{iv}+2\al_3\mu^2\phi^{\mu-1}\phi'\phi'''-
2\al_3\mu d\phi^\mu\phi''- 2d\al_3\mu^2\phi^{\mu-1}(\phi')^2+\\
+\al_2\phi'-2\al_3\mu\la\phi^{\mu+1}-2\al_3\phi=0.
\ea \ee \\
For $\alpha_2\ne 0$, the solutions $u(x,t)$ are travelling waves whose speed decreases in proportion to $t^{-1}$ and whose amplitude decreases in proportion to $t^{-1/\mu}$.

Equation (\ref{3-21}) is not integrable  but we were  able to find the   particular solutions
if $\mu=1$, \, $\al_2=0$:
\be\label{3-22} \ba{l}
\medskip
\phi(\om)=-\frac{2}{3\la}+\frac{2}{3|\la|}\sin(\ta\om+\ta_0),
\quad \ta^4+d\ta^2-\frac{\la}{2}=0 \ea \ee
and
\be\label{3-25} \ba{l}
\medskip
\phi(\om)=-\frac{2}{3\la}+C_1e^{\ta\om}+\frac{1}{9C_1\la^2}e^{-\ta\om},
\quad \ta^4-d\ta^2-\frac{\la}{2}=0. \ea \ee
Thus, using ansatz  (\ref{3-19}), we arrive at the solutions
\be\label{3-24} \ba{l}
\medskip
u=\frac{-\frac{2}{3\la}+\frac{2}{3|\la|}\sin(\ta x+\ta_0)}{t},\\
v=\frac{-\frac{2}{3|\la|}\ta^2\sin(\ta x+\ta_0)}{t}, \ea \ee which
is a spatial sinusoid for which amplitude varies in proportion
to $1/t$, and \be\label{3-26} \ba{l}
\medskip
u=\frac{-\frac{2}{3\la}+C_1e^{\ta x}+\frac{1}{9C_1\la^2}e^{-\ta x}}{t},\\
v=\frac{C_1\ta^2e^{\ta x}+\frac{\ta^2}{9C_1\la^2}e^{-\ta x}}{t} \ea \ee
of the system
\be\label{3-23} \ba{l} \medskip
u_t = -[u v_x]_x+d[u u_x]_x+\la u^2, \\
u_{xx}-v=0. \ea \ee

 In formulae (\ref{3-24}) and (\ref{3-26}), we can also shift the time $ t$
 to $ t-t_0 $ or $ t+t_0 $ with the positive parameter $t_0 $. The first shift
 leads to
 a
 %positive-valued
  solution having  blow-up at $t=t_0$, the second leads to
 a
 % negative-valued
  solution that avoids singularity
 at $t=0$ and  tends to $0$ as $t $ approaches $\infty $. Blow-up
  and extinction  are interesting phenomena in some applications.

Finally, \textbf{case 5} $\al_3\neq0$, \quad $\ga\neq 2\mu$, \quad
$\ga\neq \mu$ leads to the similarity reduction \be\label{3-27}
\om=xt^{\frac{\mu-\ga}{2(\ga-2\mu)}}, \quad
u=\phi(\om)t^{\frac{1}{\ga-2\mu}}, \quad
v=\psi(\om)t^{\frac{1-\ga+\mu}{\ga-2\mu}}\ee and to the ODE system
\be\label{3-28} \ba{l}
\medskip
(\mu-\ga)\om\phi'+2\phi=-2(\ga-2\mu)\phi^{\ga-1}[\ga\phi'\psi'+\phi\psi'']+
2(\ga-2\mu)d\phi^{\mu-1}[\mu(\phi')^2+\phi\phi'']+\\
+2(\ga-2\mu)\la\phi^{1-\ga+2\mu},\\
\psi=\phi''.
\ea \ee \\
The equivalent  4-th order equation has the form

\be\label{3-29} \ba{l}
\medskip
2(\ga-2\mu)\phi^\ga\phi^{iv}+2(\ga-2\mu)\ga
\phi^{\ga-1}\phi'\phi'''-2(\ga-2\mu)d\phi^\mu\phi''-
2d(\ga-2\mu)\mu\phi^{\mu-1}(\phi')^2+\\
+(\mu-\ga)\om\phi'-2(\ga-2\mu)\la\phi^{1-\ga+2\mu}+2\phi=0. \ea
\ee whose solutions are self-similar by a scaling invariance. Although this equation is again not integrable, one may try
to construct particular  solutions in the form of a high-order
polynomial. For example, setting  $d=0$, \quad $\ga=1$, \quad
$\mu=0$, the exact solution
\be\label{3-30} \phi(\om)=\frac{1}{120}\om^4+\frac{5}{6}\la\\
\ee
is obtained. Thus, we arrive at the solution
 \be\label{3-32} \ba{l}
\medskip
u=\frac{1}{120}\frac{x^4}{t}+\frac{5}{6}\la t,\\
v=\frac{1}{10}\frac{x^2}{t} \ea \ee of the system \be\label{3-31}
\ba{l}
\medskip
u_t = -[u v_x]_x+\la, \\
u_{xx}-v=0. \ea \ee

\noindent \textbf{Remark 2}. Using the ad hoc ansatz
\be\label{3-32a} \ba{l}
\medskip
u=\phi_0(t)+\phi_1(t)x+\phi_2(t)x^2+\phi_3(t)x^3+\phi_4(t)x^4, \\
v= 2\phi_2(t)+6\phi_3(t)x+12\phi_4(t)x^2, \ea\ee solution
(\ref{3-32}) can be generalized to the form \be\label{3-32b} \ba{l}
u=\frac{C_0}{t^{\frac{1}{5}}}+30\frac{C_1C_3}{t^{\frac{2}{5}}}+
900\frac{C_2C_3^2}{t^{\frac{3}{5}}}+
6750\frac{C_3^4}{t}+\frac{5}{6}\la t+\\ \medskip
+\Bigl(\frac{C_1}{t^{\frac{2}{5}}}+60\frac{C_2C_3}{t^{\frac{3}{5}}}+
900\frac{C_3^3}{t}\Bigr)x+\Bigl(\frac{C_2}{t^{\frac{3}{5}}}+45\frac{C_3^2}{t}\Bigr)x^2+
\frac{C_3}{t}x^3+\frac{1}{120}\frac{1}{t}x^4,\\
v= 2\Bigl(\frac{C_2}{t^{\frac{3}{5}}}+45\frac{C_3^2}{t}\Bigr)x^2+
\frac{6C_3}{t}x^2+\frac{1}{10}\frac{1}{t}x^3, \ea \ee where $C_i \,
(i=0,...,3)$ are arbitrary constants.

\bigskip

Similarly, setting  $d=0$, \quad $\ga=1$, \quad $\mu=\frac{1}{4}$,
the solution
\be\label{3-33}
\phi(\om)=\Bigl(\frac{1}{\sqrt{120}}\om^2+\frac{5}{11}\la\Bigr)^2
\ee
can be derived. Thus, the system

\be\label{3-34} \ba{l} \medskip
u_t = -[u v_x]_x+\la \sqrt{u}, \\
u_{xx}-v=0 \ea \ee

possesses the exact solution
\be\label{3-35} \ba{l}
\medskip
u=\Bigl(\frac{1}{\sqrt{120}}\frac{x^2}{\sqrt{t}}+\frac{5}{11}\la t \Bigr)^2,\\
v=\frac{1}{10}\frac{x^2}{t}+\frac{10}{11\sqrt{30}}\la\sqrt{t}. \ea
\ee

We note that solutions (\ref{3-32}) and (\ref{3-35}) have been
earlier obtained in \cite{gal-svi-07} via the method of invariant
subspaces. Indeed, the formulas (3.29) and (3.76) \cite{gal-svi-07}
contain
 (\ref{3-32}) and (\ref{3-35}), respectively.

% *********************************the case 6 of table 1***********************************

\medskip

 \begin{center}
\textbf{4. Conclusions}
\end{center}
\medskip

We have carried out a symmetry group classification for fourth-order
reaction-diffusion equations, allowing for both second-order and
fourth-order diffusion terms.  The fourth order equation has been
treated, firstly,  as a system of second-order equations that bears
some resemblance to a system of coupled reaction-diffusion equations
with cross diffusion, secondly, as a system of a second-order
equation and two first-order equations. It turns out that both
systems lead to the same result of  symmetry group classification.
Our paper generalizes  the results of Lie symmetry analysis derived
earlier for particular cases of Eq.  (\ref{1-1}). Moreover, we were
able to construct  all possible Lie symmetries, which Eq.
(\ref{1-1})
 can admit depending on the function triplets $(K, D, F)$. This
 distinguishes our investigation from those that have focussed on Lie symmetry of particular cases
 of the given fourth-order
evolution  equation. On the other hand, our result
is analogous to that derived in \cite{dor}
 and  \cite {ch-se-06} for  second-order
evolution  equations.

To the best of our knowledge, there is only the recently published
book \cite{gal-svi-07}, where exact solutions have been found for
some equations of the form (\ref{1-1}) with  $F\not=0$. Thus,  a
fourth-order nonlinear equation with the non-zero reaction term in
the form of system (\ref{3-1}) was examined by applying the Lie
symmetry reduction.
  Where possible, we have constructed exact solutions to the ordinary
  differential equations that were obtained from this reaction-diffusion
  system.
  The solutions include some unusual structures as well as the familiar types that regularly occur
  in symmetry reductions, namely self-similar solutions, decelerating and decaying traveling waves,
  and steady states. Many of the functional relationships between the two symmetry invariants are
  quite simple, involving polynomials, algebraic functions, logarithms, exponentials and sinusoids.
However, there are some that have been reduced only to the solutions
of transcendental equations (see
  formulas (\ref{3-13}) and (\ref{3-13*})).

   Finally, it should be also noted
  that the nonlinear fourth order ODEs obtained  in section 3 can be
  solved  by numerical methods and therefore solutions of the
  relevant generalized thin film equations will be constructed.


\begin{thebibliography}{99}
\footnotesize  %@@

\bibitem{Brown} Brown K J, Lacey A A (1990) Reaction--Diffusion  Equations. Oxford Univ. Press

\bibitem{Britton} Britton N F (1986)Reaction--Diffusion Equations and their Applications to Biology. Academic Press, New York

\bibitem{Cantrell} Cantrell R S, Cosner C (2003) Spatial Ecology via Reaction-Diffusion
Equations. Wiley Interscience


\bibitem{Hajek} Bradshaw--Hajek B, Broadbridge P (2004) A robust cubic  reaction--diffusion
system for gene propagation. Math  Computer Modelling 39:
1151---1163

\bibitem{Landman} Landman K A, Simpson M J, Newgreen D F (2007) Mathematical and
experimental insights into the development of the enteric nervous
system and Hirschsprung's Disease. Development, Growth and
Differentiation 49(4): 277---286

\bibitem {bertozzi-98} Bertozzi, Andrea L (1998)The mathematics of moving contact lines in thin liquid films.
Notices Amer. Math. Soc.  45(6): 689---697

\bibitem {bertozzi-pugh-98} Bertozzi A L, Pugh M C (1998) Long--wave instabilities and saturation in thin film equations. Comm. Pure Appl.
Math. 51:  625---661

\bibitem{king-et-all-07} Evans  J D, Galaktionov V A, King J R (2007) Blow--up similarity solutions of
the fourth--order unstable thin film equation.  European J. Appl.
Math.  18(2): 195---231

\bibitem {greenspan-1978} Greenspan HP. J  (1978) Fluid Mech
On the motion of a small viscous droplet that wets a surface. J
Fluid Mech  84(1): 125---143

\bibitem{Oron} Oron A, Davis S H, Bankoff S G (1997) Long--scale evolution of thin liquid films.
Rev. Mod. Phys. 69(3): 931---980

\bibitem{Dupont} Dupont T F, Goldstein R E, Kadanoff L P, Zhou S--M (1993) Finite--time singularity formation
in Hele--Shaw systems. Phys. Rev. E 47: 4182---4196

\bibitem{CH} Cahn J W, Hilliard J E (1958) Free energy of a nonuniform system.
I. Interfacial energy. J. Chem. Phys 28

\bibitem {cohen-99} Grinfeld M, Novick--Cohen A (1999)  The viscous Cahn--Hilliard equation:
Morse decomposition and structure of the global attractor.  Trans.
Amer. Math. Soc.  351(6): 2375---2406

\bibitem{Mullins}
 Mullins W (1957) Theory of thermal grooving. J. Appl. Phys. 28(3): 333---339

\bibitem{Cahn} Cahn J, Taylor J (1994) Surface motion by surface diffusion. Acta Metall. Mater. 42: 1045---1063

\bibitem{Yarin} Yarin A L, Oron A, Rosenau P (1993) Capillary instability of thin liquid film on a cylinder.
Phys. Fluids A 5: 91---98

\bibitem{Smyth}
 Smyth N F,  Hill J M (1988) High--order nonlinear diffusion. IMA J. Appl. Math. 40(2): 73---86

 \bibitem {bernis-91} Bernis F, McLeod J B (1991) Similarity solutions of a higher order nonlinear diffusion
equation. Nonl. Anal., TMA 17: 1039--–1068


\bibitem {choudhury-95} Choudhury S Roy  (1995) General similarity reductions of a family
of Cahn--Hilliard equations.  Nonlinear Anal.  24(2): 131---146

\bibitem{king-2000}  Bernis F, Hulshof J, King J (2000) Dipoles and similarity solutions
of the thin film equation in the half--line. Nonlinearity 13:
413---439

\bibitem {gandar-2000}
Gandarias M L, Bruz\'on M S  (2000) Symmetry analysis and
solutions for a family of Cahn-Hilliard equations. Reports on
Mathematical Phys., 46 (1-2): 89---97.

\bibitem {gandar-2002} Bruz\'on M S,  Gandarias M L, Medina E, Muriel E. (2002)
New symmetry reductions for a lubrication model, in: Ablowitz M J et
all (Eds.) Nonlinear Physics: Theory and Experiment. Vol.2. World
Scientific, Singapore, 143-148.

\bibitem{qu-06} Changzheng Qu (2006) Symmetries and solutions to the thin film equations.
J. Math. Anal. Appl.  317(2) 381---397.

\bibitem {gan-ibr-2008}
Gandarias M L, Ibragimov N.H.  (2008) Equivalence group of a
fourth-order evolution equation unifying various non-linear models.
Comm. in Nonlin. Sci. and Num. Simulation 13: 259–268.




 \bibitem {gal-svi-07} Galaktionov Victor A, Svirshchevskii Sergey R (2007) Exact solutions and invariant
  subspaces of nonlinear
  partial differential equations in mechanics and physics.
   Chapman \& Hall/CRC Applied Mathematics and Nonlinear Science Series.
   Chapman \& Hall/CRC, Boca Raton, FL : xxx+498 pp.

\bibitem {segel-84} Novick--Cohen Amy, Segel Lee A (1984) Nonlinear aspects of the Cahn-Hilliard equation.
Phys. D  10(3): 277---298

\bibitem {evans-gal-06} Evans J D, Galaktionov V A, Williams J F (2006) Blow--up and global
asymptotics of the limit unstable Cahn-Hilliard equation.  SIAM J.
Math. Anal.  38(1): 64---102

\bibitem{Valbusa}
Valbusa U, Boragno C, Buatier de Mongeot F (2002) Nanostructuring
surfaces by ion sputtering, J. Phys.: Condens. Matter 14:
8153---8175

\bibitem{Gaskell} Gaskell P H, Jimack P  K, Sellier  M, Thompson  H  M (2006) Flow of evaporating,
gravity-driven thin liquid films over topography. Physics of
Fluids 18(1): 013601.1---013601.14




\bibitem{ovs-59} Ovsyannikov L V (1959) Group relations of the equation of non--linear heat conductivity,
  Dokl. Akad. Nauk SSSR  125: 492---495

\bibitem {dor} Dorodnitsyn V A (1982)
 On invariant solutions of non--linear heat conduction with a source,
 USSR Comput. Math. and Math. Phys. 22: 115---122


 \bibitem{ch-se-06} Cherniha R, Serov M I (2006) Symmetries, Ansaetze  and Exact Solutions of
 Nonlinear Second--order Evolution Equations with Convection Terms. II.
  Euro.\ J.\ Appl.\ Math.\ 17: 597---605

 \bibitem {ch-se-ra-08} Cherniha R, Serov M, Rassokha I (2008) Lie Symmetries   and
Form--preserving  Transformations of
Reaction--Diffusion--Convection Equations. J. Math. Anal. Appl.
342: 1363---1379

\bibitem{Arrigo} Arrigo D J, Hill J M, Broadbridge P (1994) Nonclassical Symmetry Reductions of the
Linear Diffusion Equation with a Nonlinear Source,    I.M.A. J.
Appl. Math. 52: 1---24

\bibitem{Clarkson} Clarkson P A, Mansfield E L (1994) Symmetry reductions and exact solutions of a class
of nonlinear heat equations.Physica D70: 250---288

\bibitem {a-h} Arrigo D J, Hill J M (1995)
 Nonclassical symmetries for nonlinear diffusion and absorption
Stud. Appl.Math. 94: 21---39

\bibitem{ovs}  Ovsiannikov L V  (1980) The Group Analysis of
Differential Equations. Academic Press, New--York

 \bibitem {ch-king} Cherniha R M, King J R (2000)
Lie    Symmetries of Nonlinear  Multidimensional
Reaction--Diffusion Systems: I.
  J.    Phys.   A:      Math.   Gen. 33: 267---282, 7839---41

\bibitem {ch-king2} Cherniha R M, King J R (2003)
Lie    Symmetries of Nonlinear  Multidimensional
Reaction-Diffusion Systems: II. J. Phys. A: Math.Gen
  36: 405---425


 \bibitem {ch-king5}  Cherniha R M, King J R (2006) Lie    Symmetries and Conservation Laws  of Nonlinear
 Multidimensional Reaction--Diffusion Systems  with Variable
 Diffusivities.
  IMA J. Appl. Math. 71: 391---408

\bibitem {PB} Broadbridge P (2008)
Entropy Diagnostics for Fourth Order Partial Differential
Equations in Conservation Form. Entropy 8: 295---311

\bibitem{Ibragimov} Anderson R  L, Ibragimov, N H (1979) Lie--B\"acklund Transformations in
Applications. SIAM, Philadelphia

\bibitem{DimSYM}
Sherring J, Head A K, Prince G E (1997) Dimsym and LIE  Symmetry
Determination Packages. Mathematical and Computer Modelling, 25
(8): 153---164

\bibitem{Vu}
Vu K T, Butcher J, Carminati J  (2007) Similarity solutions of
partial differential equations using DESOLV. Computer Physics
Communications 176 (11--12): 682---693

\bibitem{CRACK}
T. Wolf (2002), Crack, LiePDE, ApplySym and ConLaw, section 4.3.5
and computer program on CD-ROM in: Grabmeier, J., Kaltofen, E. and
Weispfenning, V. (Eds.): Computer Algebra Handbook, Springer, pp.
 465---468

\bibitem {bl-co} Bluman G W, Cole J D  (1974) Similarity Methods for Differential
Equations. Springer, Berlin


\bibitem {ol} Olver P  (1986) Applications of Lie Groups to Differential
Equations. Springer, Berlin

  \bibitem{fss} Fushchych W I, Shtelen W M, Serov  M I (1993) Symmetry analysis
  and exact solutions of equations of
nonlinear mathematical physics. Kluwer, Dordrecht

\bibitem {pol-za} Polyanin A D, Zaitsev V F (2003) Handbook of
exact solutions for ordinary differential equations. CRC Press
Company.

%\bibitem {bernis-92}  Bernis F, Peletier L A, Williams S M (1992)
%Source type solutions of a fourth order nonlinear degenerate
%parabolic equation. Nonlinear Anal. 18: 217---234





















































\end{thebibliography}
\end{document}